 \documentclass[final,3p,times]{elsarticle}
\usepackage{amssymb}
\usepackage{tabularx}
\usepackage{color}
\usepackage{ulem}

\renewcommand{\theaffn}{\arabic{affn}}
\makeatletter
\long\def\@@address[#1]#2{\g@addto@macro\elsaddress{%
    \def\baselinestretch{1}%
     \refstepcounter{affn}
     \xdef\@currentlabel{\theaffn}
     \elsLabel{#1}%
    \textsuperscript{\theaffn}#2\par}}
\def\ps@pprintTitle{%
  \let\@oddhead\@empty
  \let\@evenhead\@empty
  \def\@oddfoot{\reset@font\hfil\thepage\hfil}
  \let\@evenfoot\@oddfoot
}
\makeatother

\begin{document}

\begin{frontmatter}



\title{Thermal modification of bottomonium spectra from QCD sum rules with the maximum entropy method}


\author[label1]{Kei Suzuki}\ead{k.suzuki.2010@th.phys.titech.ac.jp}
\author[label2]{Philipp Gubler} 
\author[label3]{Kenji Morita}
\author[label1,label4]{and Makoto Oka}

\address[label1]{Department of Physics, H-27, Tokyo Institute of Technology, Meguro, Tokyo 152-8551, Japan}
\address[label2]{RIKEN Nishina Center, RIKEN, Hirosawa 2-1, Wako, Saitama 351-0198, Japan}
\address[label3]{Yukawa Institute for Theoretical Physics, Kyoto University, Kyoto 606-8502, Japan}
\address[label4]{J-PARC Branch, KEK Theory Center, Institute of Particle and Nuclear Studies, High Energy Accelerator Research Organization (KEK), 203-1, Shirakata, Tokai, Ibaraki 319-1106, Japan}

\begin{abstract}

The bottomonium spectral functions at finite temperature are analyzed by employing QCD sum rules with the maximum entropy method. This approach enables us to extract the spectral functions without any phenomenological parametrization, and thus to visualize deformation of the spectral functions due to temperature effects estimated from quenched lattice QCD data. As a result, it is found that $\Upsilon$ and $\eta_b$ survive in hot matter of temperature up to at least $2.3T_c$ and $2.1T_c$, respectively, while $\chi_{b0}$ and $\chi_{b1}$ will disappear at $T<2.5T_c$. Furthermore, a detailed analysis of the vector channel shows that the spectral function in the region of the lowest peak at $T=0$ contains contributions from the excited states, $\Upsilon(2S)$ and $\Upsilon(3S)$, as well as the ground states $\Upsilon (1S)$. Our results at finite $T$ are consistent with the picture that the excited states of bottomonia dissociate at lower temperature than that of the ground state. Assuming this picture, we find that $\Upsilon(2S)$ and $\Upsilon(3S)$ disappear at $T = 1.5 -2.0T_c$.
\end{abstract}


\end{frontmatter}


\section{Introduction}  \label{Introduction}
Properties of high density and temperature matter is one of the most exciting subjects of hadron physics. Quantum chromodynamics (QCD) predicts matter composed of quarks and gluons in the form of unconfined plasma phase, called quark gluon plasma (QGP). In such a phase of matter, production of quarkonium, a bound state of a heavy quark and a heavy antiquark, is expected to be suppressed. Especially, $J/\psi$ suppression has traditionally been considered as a signal for the formation of QGP \cite{Matsui1986,Hashimoto1986} in high-energy heavy ion collisions. There, the charmonia including $J/\psi$ may dissociate in the QGP due to temperature effects such as the color Debye screening. Such suppressions were observed in the heavy ion collisions at the Relativistic Heavy Ion Collider (RHIC). A similar phenomenon may occur in the bottomonium spectrum. Recent report \cite{Chatrchyan2011} has suggested significant modification of the bottomonium spectra from the comparison between the P-P and Pb-Pb collisions at the Large Hadron Collider (LHC). The data indicate larger suppression of the excited states than the ground state. One purpose of the present paper is to study this phenomenon by using QCD sum rules with MEM. It is thus interesting to see whether there are qualitative/quantitative similarities and differences between the behavior of the charmonia and bottomonia spectra at finite temperature.

In order to quantify and predict when and how the quarkonium spectrum is modified, various theoretical approaches have been developed. They include Lattice QCD \cite{Umeda2001,Asakawa2004,Datta2004,Umeda2005,Jakovac2007,Aarts2007,Ding2010,Rothkopf2012,Aarts2011,Karsch2012}, QCD sum rules \cite{Morita2008,Morita2008-2,Song2009,Morita2010,Gubler2011}, AdS/QCD \cite{Kim2007,Fujita2009,Noronha2009,Grigoryan2010}, resummed perturbation theory \cite{Laine2007, Laine2007-2}, and effective field theories \cite{Brambilla2008, Brambilla2010} as well as potential models \cite{Rothkopf2012,Wong2005, Digal2005, Satz2006, Alberico2005, Mocsy2008, Mocsy2007, Petreczky2011, Cabrera2007, Riek2010, Riek2011}. Most of the previous studies were devoted to the charmonium spectra, while, for bottomonia, there are only a few theoretical results \cite{Aarts2011, Morita2010, Wong2005, Satz2006, Petreczky2011, Cabrera2007, Riek2010, Strickland2012}, and it is important that they are examined by independent methods.  

In a previous work \cite{Gubler2011}, we have investigated the behavior of charmonium spectral functions at finite temperature from QCD sum rules with the maximum entropy method (MEM). Our approach enables us to extract directly the shape of the spectral functions. Therefore, this is a suitable tool to study the deformation of the spectral functions upon the change of the temperature. In this paper, we will study the bottomonium spectrum by using the same method. We point out that the bottomonium spectral function is crucially different from that of charmonium in the sense that there are several excited states below the continuum threshold. For example, $\Upsilon$ has the ground state ($1S$) and the excited states ($2S$, $3S$).

This paper is organized as follows: In Section \ref{Analysis procedure}, we discuss QCD sum rules for heavy quarkonia at finite temperature and demonstrate our method to extract spectral functions with MEM. In Section \ref{Results and discussion}, we show the obtained spectral functions at zero and finite temperature. Here, we also investigate the behavior of the excited states of the bottomonium vector channel. Section \ref{Conclusion} is devoted to the summary and conclusion. 

\section{Analysis procedure}  \label{Analysis procedure}
This section overviews our analysis method of QCD sum rules with MEM. Historically, QCD sum rules for heavy quarkonia were introduced in \cite{Shifman1979, Shifman1979-2} and then elaborated in \cite{Reinders1981}. Their Borel transformation were calculated in \cite{Bertlmann1982}. Applying MEM to QCD sum rules was successfully performed for the $\rho$ meson in \cite{Gubler2010} and the nucleon in \cite{Ohtani2011, Ohtani2012}.  We follow the same procedure as \cite{Gubler2011} for the QCD sum rules of heavy quarkonia at finite temperature.

\subsection{QCD sum rules for bottomonia}
Let us start with the current correlation function
\begin{equation}
\Pi^{\, J}(q) = i \int d^4x e^{i q \cdot x} \langle T[j^{\, J}(x) j^{\, J \dag}(0)] \rangle , \label{correlation function}
\end{equation}
where $J$ stands for the pseudoscalar $(P)$, vector$(V)$, scalar $(S)$, and axial-vector $(A)$ channel. Each current is defined as $j^{\, P}=\bar{b} \gamma_5 b$, $ j^{\, V}_\mu = \bar{b} \gamma_\mu b$, $ j^{\, S} = \bar{b}b$, and $ j^{\, A}_\mu = (q_\mu q_\nu /q^2 -g_{\mu\nu})\bar{b} \gamma_5 \gamma^\nu b$ with $b$ being the bottom quark operator. For the axial-vector current, the projection to the transverse components for the $\chi_{b1}$ states is carried out. At finite temperature, the currents of the $V$ and the $A$ channels generally have two independent components. We assume spatial momentum of the bottomonia to be zero so that only one component becomes independent : $q^\mu=(\omega, \bf{0})$. Then, one can define the dimensionless correlation functions as $\tilde{\Pi}^{P, \, S}(q^2) = \Pi^{P, \, S}(q)/q^2 $ and $\tilde{\Pi}^{V, \, A}(q^2) = \Pi_\mu^{\mu V, \, A}(q)/(-3q^2) $.

Using the operator product expansion (OPE), one can expand the operator $j^{\, J}(x) j^{\, J \dag}(0)$ of Eq.(\ref{correlation function}) as a series of local operators $O_n$ with mass dimension $n$. Then, the dimensionless correlation functions are given as 
\begin{equation}
\tilde{\Pi}^{\, J}(q^2) = \sum_n C_n^{\, J} (q^2) \langle O_n \rangle. \label{correlation function2}
\end{equation}
If the scale of the gluon condensates is smaller than the separation scale: $4m_b^2-q^2 \gg  \langle G \rangle \sim (\Lambda_{\mathrm QCD} +aT + b \mu)^2$, one can assume that all the temperature effects are included in the expectation values of the local operators $ \langle O_n \rangle$ \cite{Hatsuda1993,Morita2010}. Thus the Wilson coefficients $ C_n^{\, J} (q^2)$ can be considered to be independent of $T$, as long as the temperature is not too high.

In order to improve the OPE convergence and suppress contributions of higher energy states, we perform the Borel transformation on Eq.(\ref{correlation function2}). Then, the correlation function can be written down as
\begin{equation}
{\mathcal M}^{\, J}(\nu) = \pi e^{-\nu} A^J(\nu) [1+\alpha_s(\nu) a^J(\nu) +b^J(\nu) \phi_b(T)+c^J(\nu) \phi_c(T)], \label{OPE}
\end{equation}
where we use the dimensionless parameter $\nu \equiv 4m_b^2/M^2$ with $M$ standing for the usual Borel mass. Now the condition for the separation scale is given by $d! M^{2d} \gg  \langle G^d \rangle$ \cite{Morita2010}. The first term is the leading order of the OPE corresponding to the free current correlation function. The second term stands for the perturbative $\alpha_s$ correction. The third and fourth terms include contributions of the scalar and twist-2 gluon condensates of mass dimension 4. Here, $\phi_b$ and $\phi_c$ are defined as
\begin{equation}
\phi_b(T) = \frac{4\pi^2}{9(4m_b^2)^2} G_0(T), \label{coefficient_G_0}
\end{equation}
\begin{equation}
\phi_c(T) = \frac{4\pi^2}{3(4m_b^2)^2} G_2(T), \label{coefficient_G_2}
\end{equation}
where $G_0(T)$ and $G_2(T)$ are the scalar and twist-2 gluon condensates at finite temperature. $G_0$ and $G_2$ are defined as $G_0(T) =  \langle \frac{\alpha_s}{\pi} G_{\mu\nu}^a G^{a\mu\nu} \rangle$ and $( u^\mu u^\nu -\frac{1}{4} g^{\mu\nu} ) G_2(T) =  \langle \frac{\alpha_s}{\pi} G_\rho^{a\mu} G^{a\nu\rho} \rangle$, where $u^\mu$ is the four velocity of the medium. In our previous work on charmonium, we have added the scalar gluon condensate of mass dimension 6 which was found to be small. Thus, we can safely neglect it because its coefficient is strongly suppressed by the bottom quark mass. The detailed expressions of the Wilson coefficients $ A^J(\nu)$, $a^J(\nu)$, $b^J(\nu)$ and $c^J(\nu)$ are given in Ref. \cite{Morita2010}.

The temperature dependences of the gluon condensates are obtained from the approach proposed in Refs. \cite{Morita2008,Morita2008-2}, where the dimension-4 gluon condensates are related to the energy-momentum tensor, which can be expressed in terms of the energy density $\epsilon$, the pressure $p$ and the strong coupling constant $\alpha_s$. In concrete, $G_0(T)=G_0^{\mathrm vac} -\frac{8}{11}[\epsilon(T) - 3p(T)]$ and $G_2(T) = -\frac{\alpha_s(T)}{\pi}[\epsilon(T)+p(T)]$. We then utilize the results of quenched lattice QCD \cite{Boyd1996,Kaczmarek2004} to obtain the temperature dependence of $\epsilon(T)$, $p(T)$ and $\alpha_s(T)$.

It should be noted here that, in the quenched approximation, the value of the critical temperature $T_c$ is about 260 MeV \cite{Boyd1996}, while the cross-over temperature of full QCD is estimated to be in the region of 145-165 MeV \cite{Borsanyi2010,Bazavov2012}\footnote{Whereas the phase transition in pure SU(3) theory is a first order one, full QCD at physical quark masses exhibits a cross-over \cite{Aoki2006}.}. In order to make the predictions more realistic, full lattice QCD with physical quark masses \cite{Borsanyi2010-2} may be applied. We, however, note that the gluon condensates are not directly extracted from energy density and pressure in full QCD because of the light quark contribution. Therefore, in the present approach, we have to assume that $G_0(T)$, $G_2(T)$ and $\alpha_s(T)$ are functions of $T/T_c$. For the scalar gluon condensate, this approximation is known to be a good around $T/T_c \sim 1.0$ \cite{Miller2006, Morita2008-2}. Thus this approach may provide a good approximation for charmonium systems. As we shall see, however, bottomonium systems receive significant modification at higher temperatures where not only $G_2(T)$, unknown for full QCD, becomes more important \cite{Morita2010} but also $G_0(T)$ in full QCD deviates from the scaling behavior. Therefore, resultant temperature dependence of the spectral functions, which will be discussed below, provides no more than qualitative guides for the full QCD case. Nevertheless, we emphasize that the results obtained in this paper, rather than being entirely applicable to full QCD, can provide useful information on the difference between the behavior of the S-wave and P-wave (or charmonium and bottomonium) states.

The sum rule is constructed from the dispersion relation derived from the analytic properties of the correlation function of Eq.(\ref{correlation function}). For the spectral function $\rho^{\, J}(\omega)$, we obtain
\begin{equation}
{\mathcal M}^{\, J}(\nu) = \int_0^\infty d\omega^2 \, e^{- \nu \omega^2 /4m_b^2} \rho^{\, J}(\omega), \label{Borel_sum_rule_initial}
\end{equation}
where the left-hand sides are equal to Eq.(\ref{OPE}). Note that, in the vector, scalar, and axial-vector channels, there is an additional constant term to Eq.(\ref{OPE}) at finite temperature, which originates from a pole at $\omega=0$ in $ \rho^{\, J}(\omega)$ and is called scattering term \cite{Bochkarev1986}. Although the contribution of this term, proportional to $e^{-m_b/T}$ , should be much smaller than the charmonium case, we differentiate Eq.(\ref{Borel_sum_rule_initial}) with respect to $\nu$ to eliminate this contribution:
\begin{equation}
\frac{\partial }{\partial \nu} {\mathcal M}^{\, J}(\nu)= -\frac{1}{4m_{b}^2} \int_0^\infty d\omega^2 \, \omega^2 e^{- \nu \omega^2 /4m_b^2} \rho^{\, J}(\omega). \label{Borel_sum_rule}
\end{equation}
To compare all the channels on the same basis, we also differentiate the sum rule of the pseudo-scalar channel. The validity of this procedure in the heavy quark sum rules was discussed in Ref. \cite{Morita2010-2}. Also, we have checked that the same results are obtained from both the original and differentiated sum rules for the pseudo-scalar channel. 

\subsection{MEM analysis of QCD sum rules}
In this subsection, we demonstrate our method to extract the spectral function $\rho^{\, J}(\omega)$ from Eq.(\ref{Borel_sum_rule}). The conventional methods of analyzing QCD sum rules assume a particular form for the spectral function, the most popular one being the ``pole + continuum'' form. By contrast, such an assumption is not necessary in our method \cite{Gubler2010} as the shape of spectral functions is directly extracted from the MEM.

Let us now briefly summarize the procedure of the MEM analysis. The basic idea is Bayes' theorem :  
\begin{equation}
P[\rho |\mathcal{M} H] =\frac{P[\mathcal{M} |\rho H] P[\rho | H]}{ P [\mathcal{M}|H]}, \label{Bayes_theorem}
\end{equation}
where $\rho$ and $\mathcal{M} $ are corresponding to the spectral function and the OPE in Eq.(\ref{Borel_sum_rule}), respectively. $H$ denotes prior knowledge on $\rho$ such as positivity and its asymptotic values. $P[\rho |\mathcal{M} H]$ represents the conditional probability of $\rho$ given $\mathcal{M}$ and $H$. On the right-hand side, $P[\mathcal{M} |\rho H]$ is called the ``likelihood function'', and $ P[\rho | H]$ stands for the ``prior probability''. $P [\mathcal{M}|H]$ is only a normalization constant and can be ignored in later discussion since it does not depend on $\rho$. In order to maximize $P[\rho |\mathcal{M} H]$, we estimate $P[\mathcal{M} |\rho H]$ and $ P[\rho | H]$.

The likelihood function is written as
\begin{equation}
P[\mathcal{M} |\rho H] = e^{-L[\rho]}, \label{likelifood_function0}\\
\end{equation}
\begin{equation}
L[\rho] = \frac{1}{2(\nu_{\mathrm{max}}-\nu_{\mathrm{min}})} \int_{\nu_{\mathrm{min}}}^{\nu_{\mathrm{max}}} d\nu \frac{[\mathcal{M} (\nu) - \mathcal{M}_\rho(\nu)]^2}{\sigma^2(\nu)}. \label{likelifood_function}
\end{equation}
Here, $\mathcal{M} (\nu)$ is obtained from the results of the OPE and corresponds to the left-hand side in Eq.(\ref{Borel_sum_rule}), while $\mathcal{M}_\rho(\nu)$ is defined as the right-hand one. $\sigma(\nu)$ stands for the uncertainty of $\mathcal{M}(\nu)$ (see Ref. \cite{Gubler2010}). On the other hand, the prior probability is written as
\begin{equation}
P[\rho | H] = e^{\alpha S[\rho]}, \label{shannon-jaynes_entropy0}\\
\end{equation}
\begin{equation}
S[\rho] = \int_0^\infty d\omega \left[ \rho(\omega) - m(\omega) - \rho(\omega) \log \left( \frac{\rho(\omega)}{m(\omega)} \right) \right], \label{shannon-jaynes_entropy}
\end{equation}
where $S[\rho]$ is known as the Shannon-Jaynes entropy and $\alpha$ is introduced as a scaling factor. $m(\omega)$ is called the default model and determines the spectral function when there is no information from the OPE. For the default model, we use a constant corresponding to the perturbative value of the spectral functions at high energy.

Using Eq.(\ref{likelifood_function0}) and  Eqs.(\ref{shannon-jaynes_entropy0}), one can rewrite Eq.(\ref{Bayes_theorem}) as
\begin{eqnarray}
P[\rho|\mathcal{M} H] &\propto& P[\mathcal{M}|\rho H] P[\rho|H] \nonumber\\
&=& e^{Q[\rho]}, \\
Q[\rho] &\equiv& \alpha S[\rho] - L[\rho].
\end{eqnarray}
In order to determine the most probable $\rho(\omega)$, we find the maximum of the functional $Q[\rho]$ by the Bryan algorithm \cite{Bryan1990}. It can be proven that the maximum of $Q[\rho]$ is unique if it exists, as it is shown explicitly in \cite{Asakawa2001}.

Before starting the MEM analysis, we have to determine the criterion for the range of the dimensionless parameter $\nu$ which is used for the investigation. We determine the upper bound $\nu_{\mathrm{max}}$ from the criterion that the perturbative $\alpha_s$ correction term contributes less than 30 \% of the leading term as in \cite{Morita2010}. The reason why we use the perturbative part rather than the term of the highest gluonic condensate as a criterion of the OPE convergence, is that for bottomonium the terms proportional to the gluon condensates are strongly suppressed due to the large bottom quark mass, as seen in Eqs.(\ref{coefficient_G_0}) and (\ref{coefficient_G_2}). Therefore, there is a region of $\nu$, where the $\alpha_s$ correction is forbiddingly large although the contributions of the condensates seem to converge. On the other hand, for the lower bound $\nu_{\mathrm{min}}$, we do not have to impose such a criterion unlike in usual QCD sum rule analyses, since we are not assuming the pole dominance of the dispersion integral. Thus, we choose  the value which obtains the highest peak as the criterion of $\nu_{\mathrm{min}}$. The range of $\nu$ (namely, Borel window) for each channel is summarized in Table \ref{Borel_window}. We have checked that the obtained results do not depend on the chosen values of these bounds. 
\begin{table}[h]
  \begin{center}
  \begin{tabular}{c|cccc}
\hline \hline
   Channel                                                & Vector ($\Upsilon$) & Pseudoscalar ($\eta_b$) & Scalar ($\chi_{b0}$) & Axial-vector ($\chi_{b1}$) \\
\hline
$\nu_{\mathrm{min}} - \nu_{\mathrm{max}}$ & 4.00-8.23              & 4.30-9.33                     & 4.00-7.33              & 3.50-5.83                        \\
\hline \hline
  \end{tabular}
  \end{center}
   \caption{Range of dimensionless parameter $\nu$.}
   \label{Borel_window}
\end{table}

\section{Results and discussion} \label{Results and discussion}

\subsection{Analysis of mock data}
In order to estimate the resolution of the MEM, we should carry out a test analysis by using mock data generated from the experimental information for the vector channel, delta functions for the bound states and a smooth step-like function for the continuum:
\begin{equation}
\mathrm{Im} \Pi^V(s)=\sum_{\mathrm{res}} \frac{27}{4} \cdot \Gamma_R (e^+ \, e^-) \frac{m_R}{\alpha^2} \cdot \delta(s - m_R^2) + \mathrm{continuum}, \label{mock_data}
\end{equation}
where, the sum is taken for the these bottomonium states, $\Upsilon(1S)$, $\Upsilon(2S)$ and $\Upsilon(3S)$. $m_R$ represents their masses, $\Gamma_R (e^+ \, e^-)$ is the corresponding decay width to an electron-positron pair, and $\alpha$ is the fine structure constant (see Ref. \cite{Reinders1981}). We construct mock data by substituting the phenomenological spectral function into Eq.(\ref{Borel_sum_rule}) and evaluating the integral numerically over $s$. We also consider the mock data for the individual bottomonium states, where a single state + continuum is taken for Eq.(\ref{mock_data}).

\begin{figure}[h!]
     \begin{center}
    \begin{minipage}[t]{0.5\columnwidth}
        \begin{center}
            \includegraphics[clip, width=1.0\columnwidth]{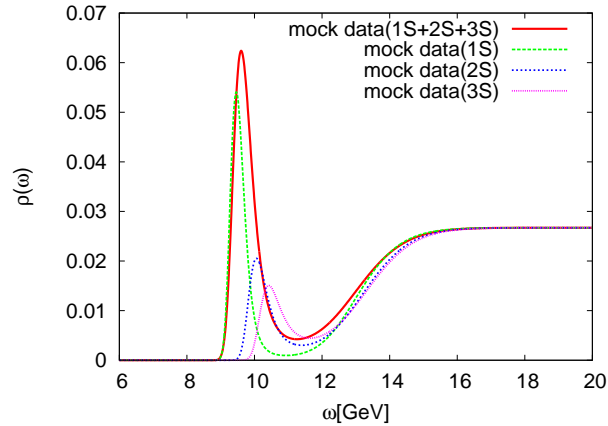}
        \end{center}
    \end{minipage}
    \end{center}
     \caption{Spectral functions extracted from mock data based on Eq.(\ref{mock_data}) with MEM.}
     \label{mockdata}
\end{figure}

The results of the mock data for the individual bottomonium states are shown in Fig. \ref{mockdata} as the dotted lines. The result for the full spectrum with all the bound states is shown in Fig. \ref{mockdata} as the solid line. One sees that the individual peaks are not resolved in the full spectrum. Namely, they are combined into a single peak after the MEM analysis. Moreover, the residue of the single peaks is consistent with the sum of the residues of $\Upsilon(1S)$, $\Upsilon(2S)$ and $\Upsilon(3S)$. It should be noted that the peak position of the combined spectrum is located at a higher energy than the $\Upsilon(1S)$ peak. The values of the peak positions and the residues are summarized in Table \ref{mocdata_table}. Later, we will use this property of the combined peak, when we extract the temperature dependence of the excited bottomonium states. 

\begin{table}[h!]
  \begin{center}
  \begin{tabular}{l|cccc}
\hline \hline
                           & $\Upsilon(1S)$ & $\Upsilon(2S)$ & $\Upsilon(3S)$ & $\Upsilon(1S+2S+3S)$ \\
\hline
mass (exp.)  [GeV]          & 9.46        & 10.02        & 10.36       &    -        \\
residue (exp.) [GeV]        & 0.0270    &  0.0123      & 0.0089     & 0.0483   \\
\hline
mass (from mock data) [GeV]   & 9.46         & 10.06        & 10.42      & 9.61     \\
residue (from mock data) [GeV]& 0.0303     & 0.0170      & 0.0147    & 0.0506 \\
\hline \hline
  \end{tabular}
  \end{center}
   \caption{Mass and residue values obtained from mock data based on Eq.(\ref{mock_data}) with MEM.}
   \label{mocdata_table}
\end{table}

\subsection{Bottomonia at zero temperature}

\begin{figure}[b!]
    \begin{minipage}[t]{0.5\columnwidth}
        \begin{center}
            \includegraphics[clip, width=1.0\columnwidth]{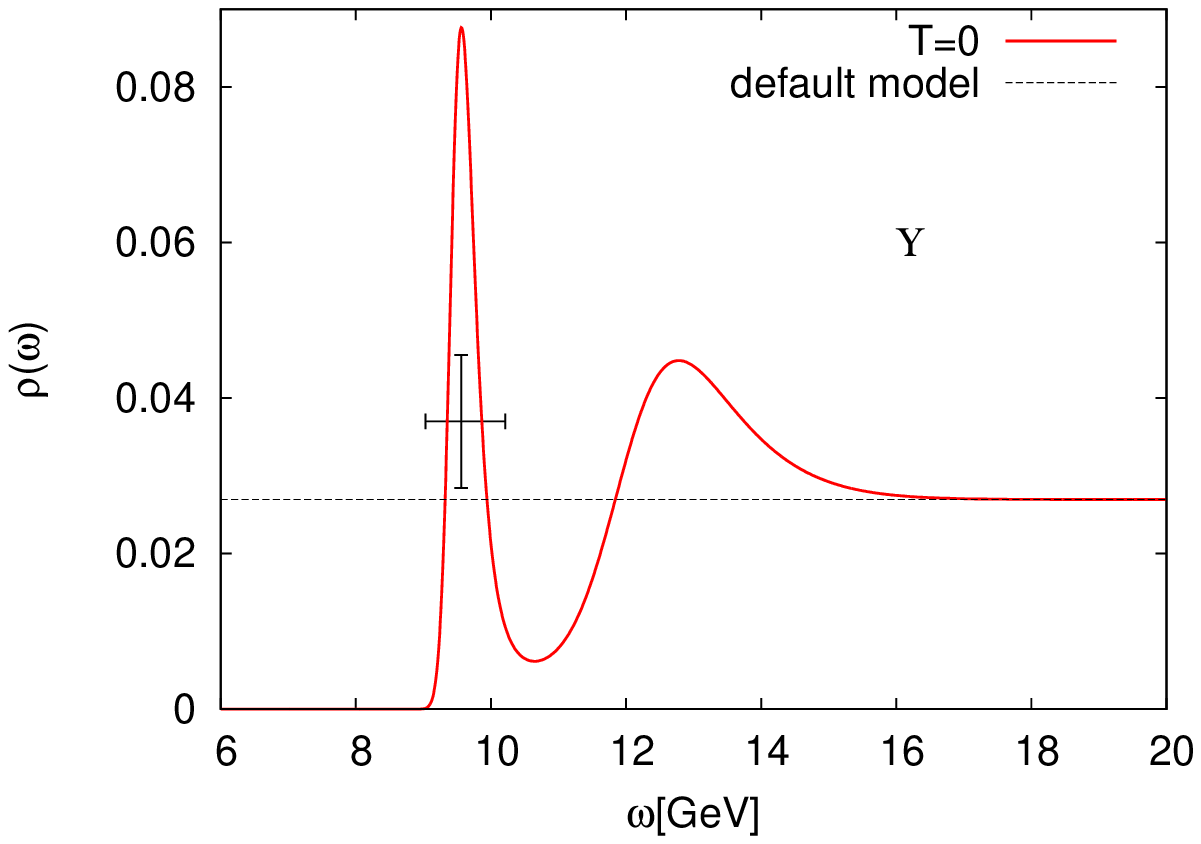}
        \end{center}
    \end{minipage}%
    \begin{minipage}[t]{0.5\columnwidth}
        \begin{center}
            \includegraphics[clip, width=1.0\columnwidth]{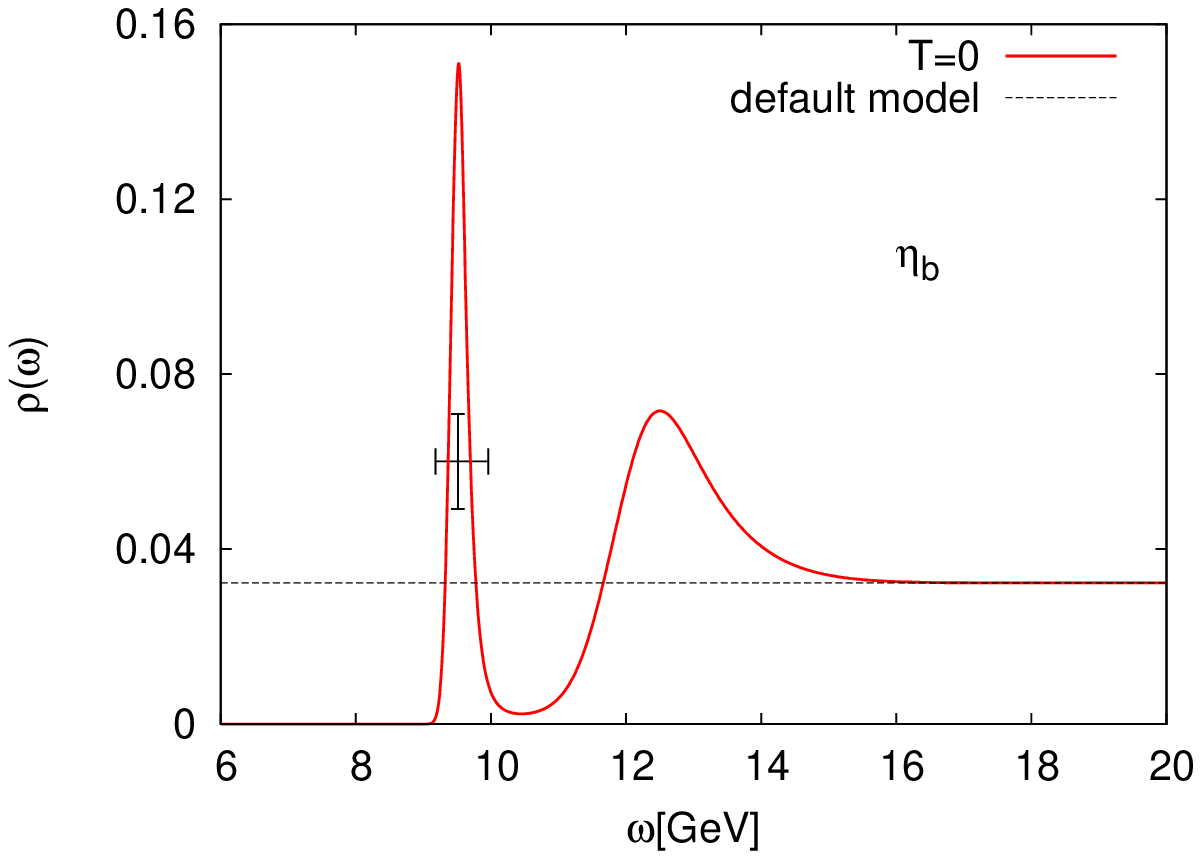}
        \end{center}
    \end{minipage}
    \begin{minipage}[t]{0.5\columnwidth}
        \begin{center}
            \includegraphics[clip, width=1.0\columnwidth]{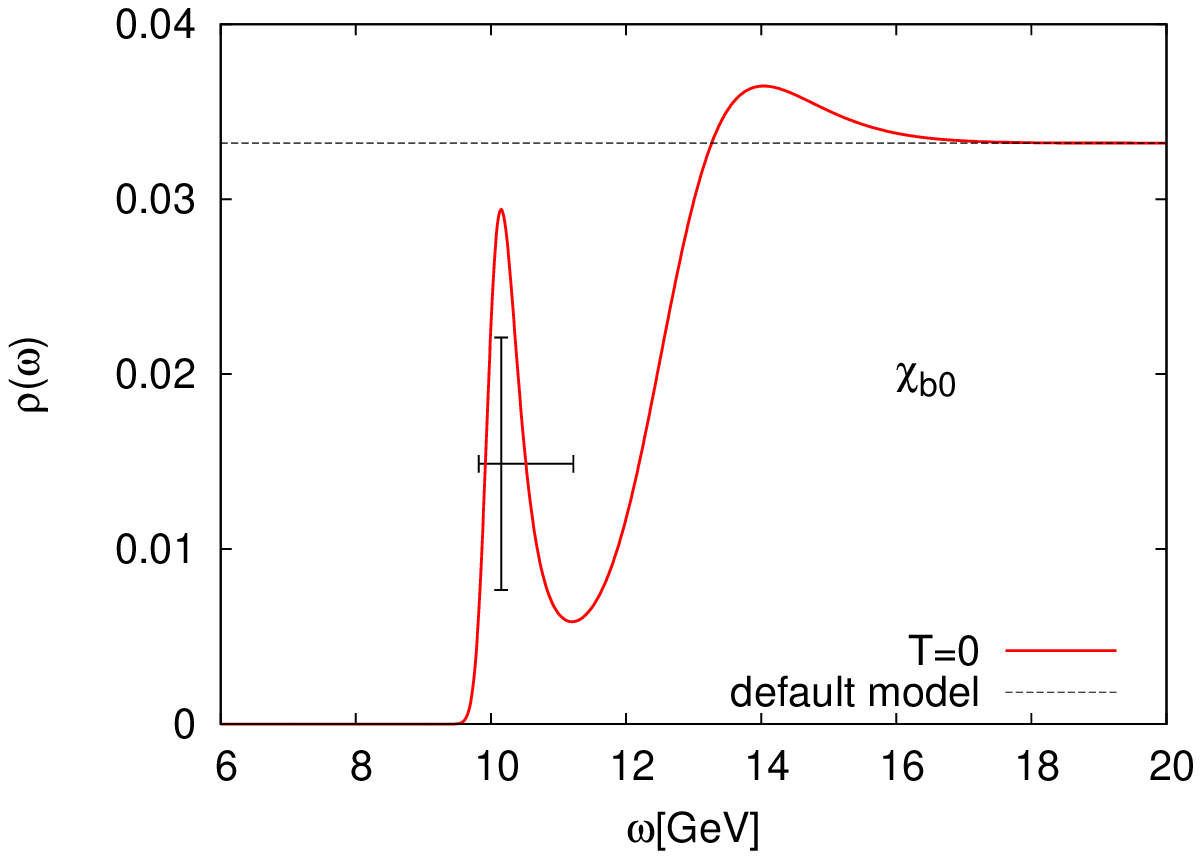}
        \end{center}
    \end{minipage}%
    \begin{minipage}[t]{0.5\columnwidth}
        \begin{center}
            \includegraphics[clip, width=1.0\columnwidth]{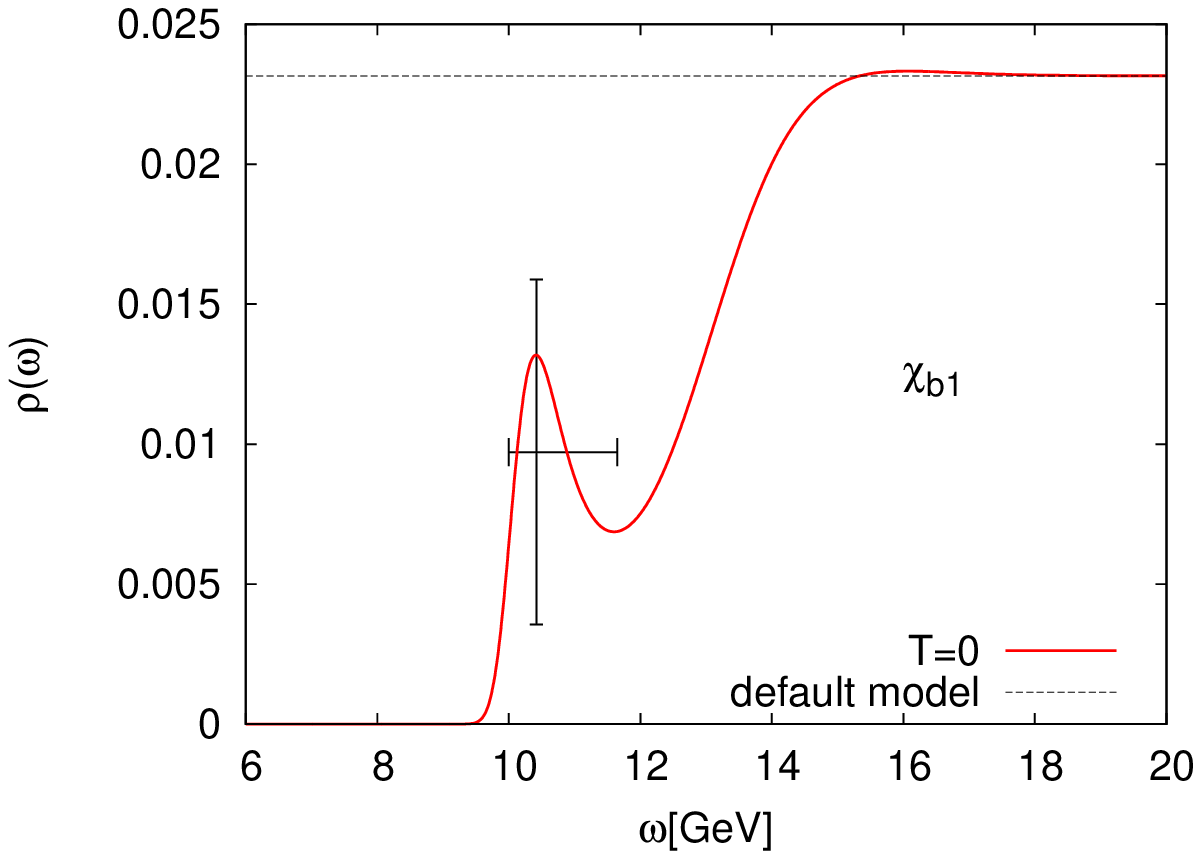}
        \end{center}
    \end{minipage}
    \caption{Spectral functions of bottomonia at zero temperature. Upper left : vector ($\Upsilon$), upper right : pseudoscalar ($\eta_b$), lower left : scalar ($\chi_{b0}$), lower right : axial-vector ($\chi_{b1}$). Vertical lines stand for statistical error of peaks obtained from MEM, while horizontal lines represent corresponding range of $\omega$, for which the error is calculated.}
    \label{figure_zero_all}
\end{figure}

The results of the MEM for the spectral functions of the vector, pseudoscalar, scalar and axial-vector $b\bar{b}$ systems at zero temperature are shown in Fig. \ref{figure_zero_all}. We employ the bottom quark mass $\bar{m}_b(m_b) = 4.167 \pm 0.013 \mathrm{GeV}$ \cite{Narison2011}, the strong coupling constant $\alpha_s(M_Z)=0.1184\pm0.0007$ \cite{Bethke2009} and the vacuum gluon condensate $G_0^\mathrm{vac} = 0.012 \pm 0.0036 \mathrm{GeV}^4$ \cite{Shifman1979}. Each spectral function shows a clear peak at around $\omega \sim 10\mathrm{GeV}$. The average peak height and the estimated error of the average height in the MEM procedure are given by the horizontal and vertical lines. The lines are drawn at the range of $\omega$ where the average is taken. One sees that the lowest-energy peaks of the S-wave channels are statistically significant. On the other hand, the lowest P-wave peaks have heights 
of the same order as their error-bars, the scalar channel being a bit higher while the axial-vector channel is lower. This means that with the precision presently available, we cannot make any strong statement about the existence of these peaks and about their behavior at finite temperature. Therefore, the results about the lowest P-wave peaks obtained in this paper should not be considered to be fully conclusive. Furthermore, for all peaks appearing at higher-energy, their statistical significances are not good. The lowest peaks are located at $m_{\Upsilon}=9.56$GeV, $m_{\eta_b}=9.51$GeV, $m_{\chi_{b0}}=10.15$GeV and $m_{\chi_{b1}}=10.42$GeV, respectively. These values are somewhat higher than the experimentally observed masses ($m_{\mathrm{exp.}}=9.460$GeV, $9.389$GeV, $9.859$GeV, and $9.893$GeV, respectively). In fact, they are consistent with our analysis of mock data since the obtained peaks contain contributions from both the ground and excited states and their positions are shifted to higher energies. Validity of this picture is confirmed by evaluating the residue of the peak for the vector channel and comparing it with the residue obtained from the leptonic decay width (the value obtained from the present analysis is 0.0476 GeV, which should be compared to Table \ref{mocdata_table}).

We have checked that such contributions of the excited states are also present in analyses by usual QCD sum rules, where the ``pole + continuum'' assumption is used, i.e. ground state cannot be separated as a single pole.

\subsection{Bottomonia at finite temperature} \label{Bottomonia at finite temperature}
\begin{figure}[b!]
    \begin{minipage}[t]{0.5\columnwidth}
        \begin{center}
            \includegraphics[clip, width=1.0\columnwidth]{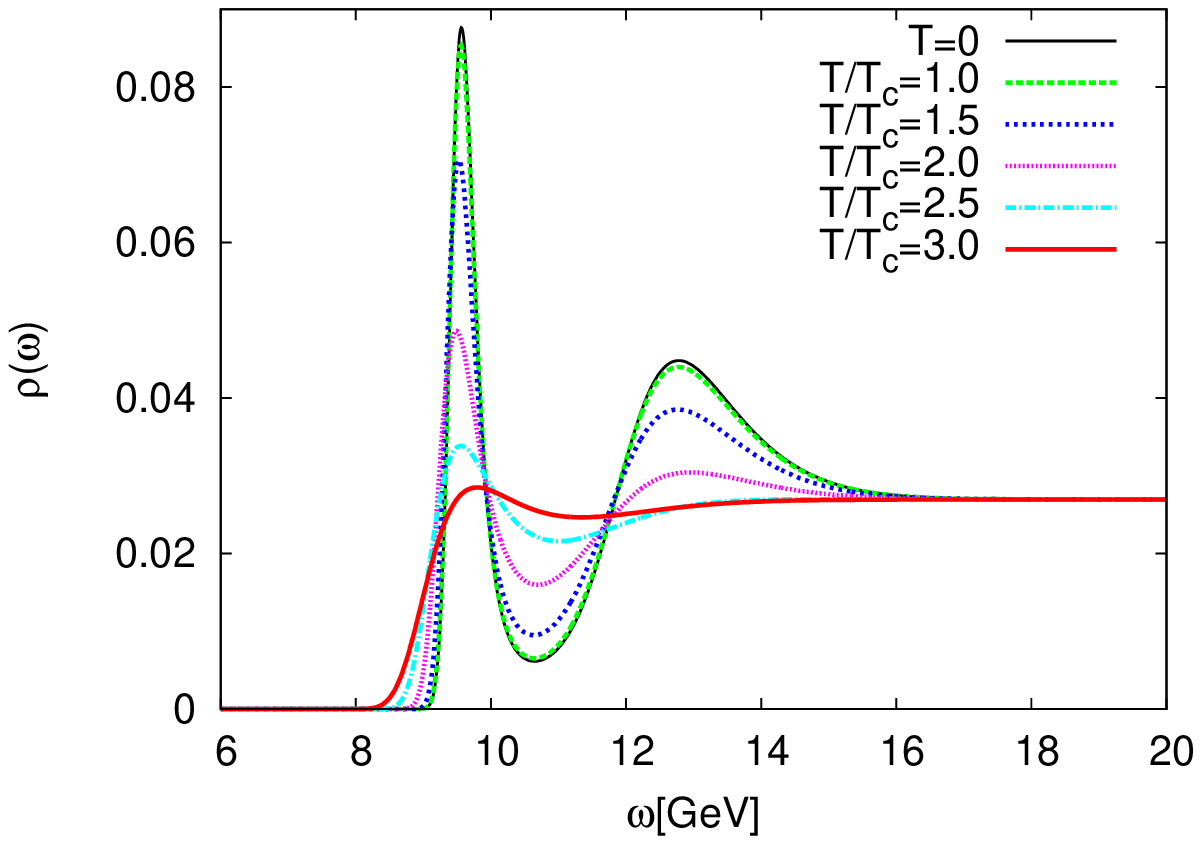}
        \end{center}
    \end{minipage}%
    \begin{minipage}[t]{0.5\columnwidth}
        \begin{center}
            \includegraphics[clip, width=1.0\columnwidth]{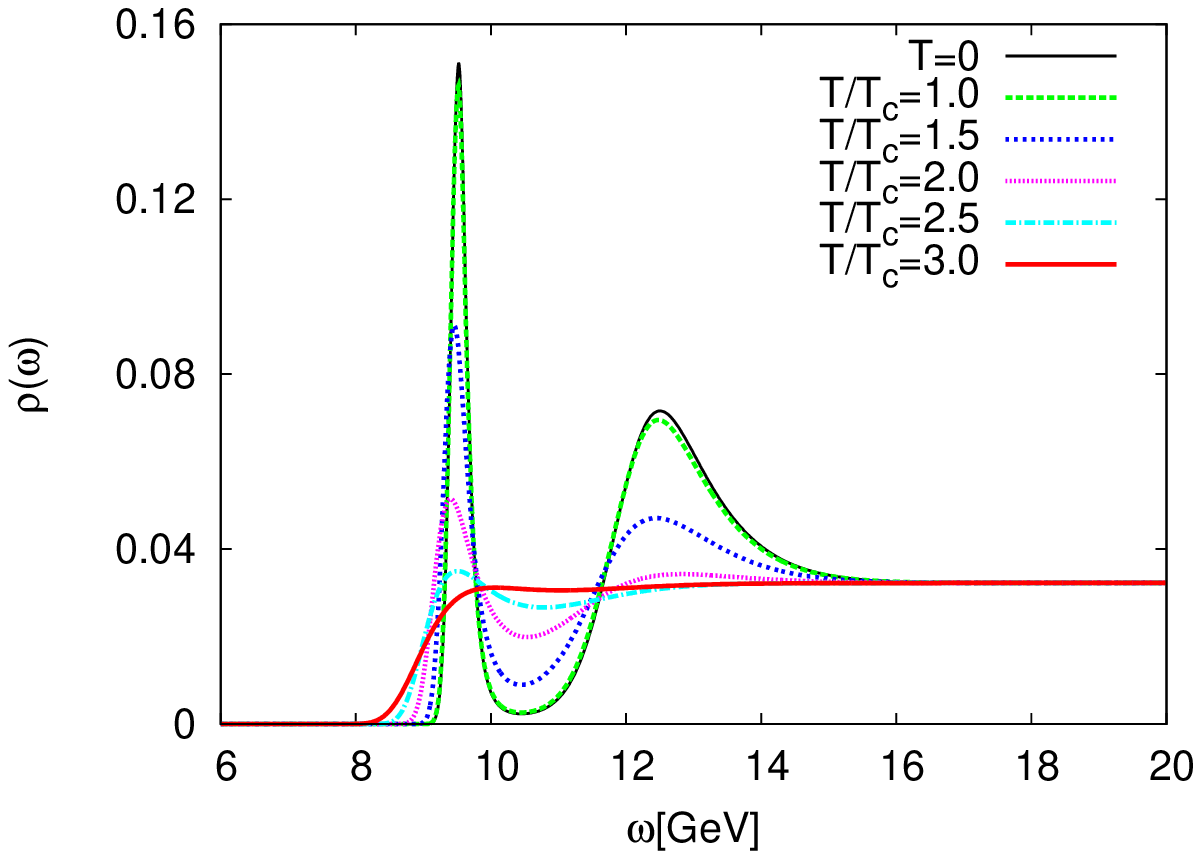}
        \end{center}
    \end{minipage}
    \begin{minipage}[t]{0.5\columnwidth}
        \begin{center}
            \includegraphics[clip, width=1.0\columnwidth]{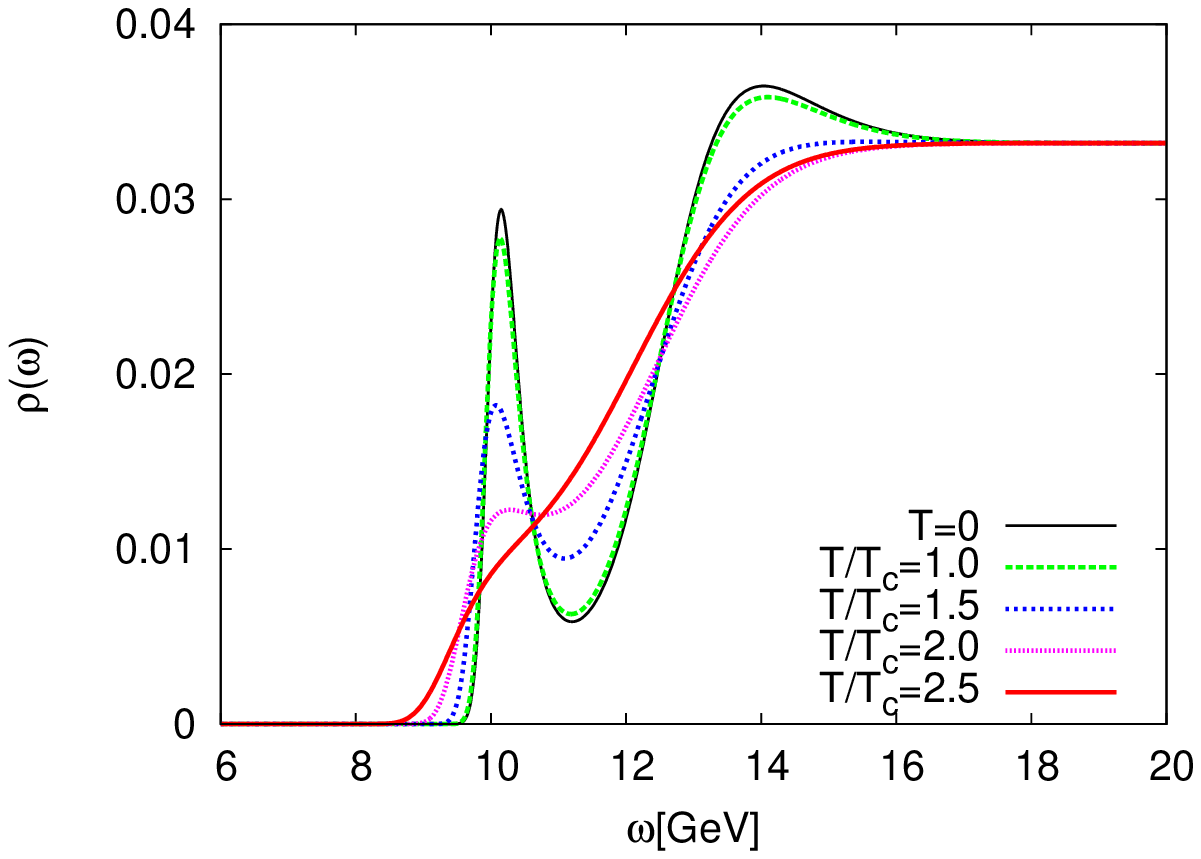}
        \end{center}
    \end{minipage}%
    \begin{minipage}[t]{0.5\columnwidth}
        \begin{center}
            \includegraphics[clip, width=1.0\columnwidth]{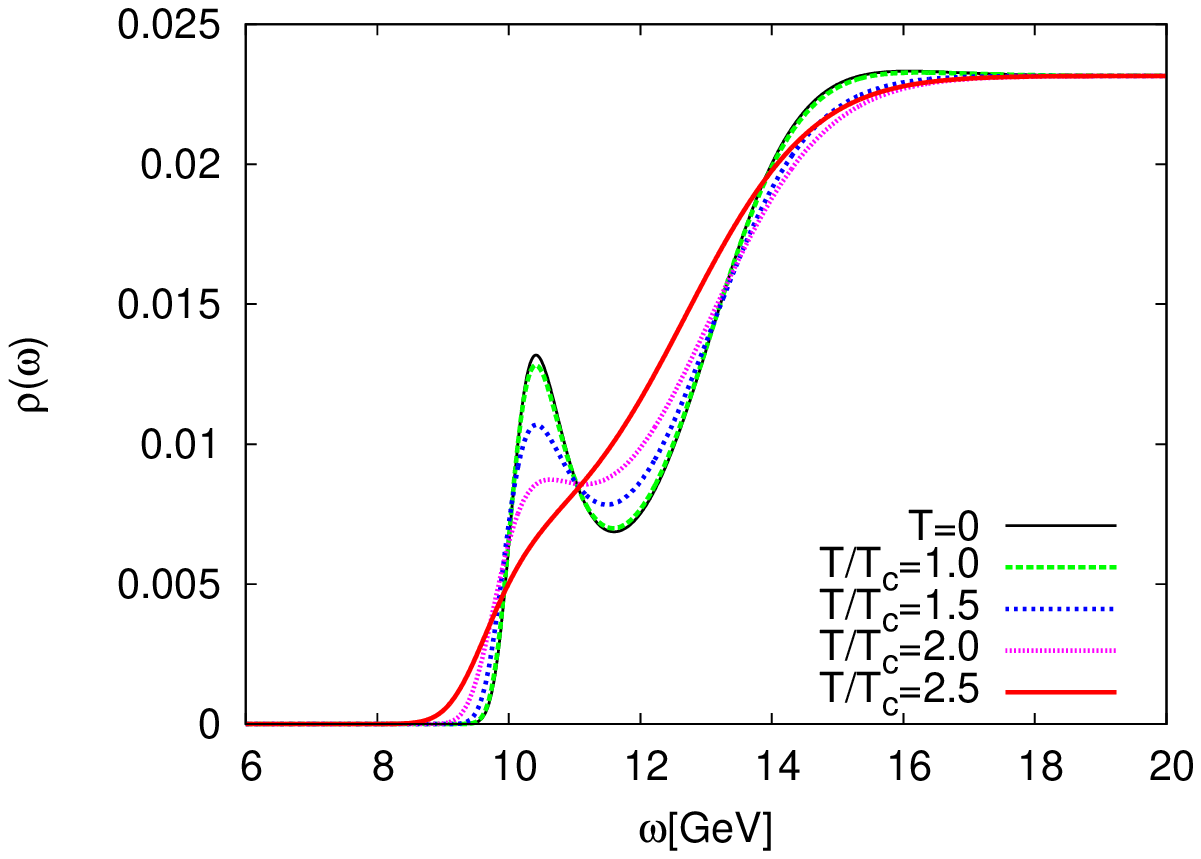}
        \end{center}
    \end{minipage}
    \caption{Spectral functions of bottomonia at finite temperature. Upper left : vector ($\Upsilon$), upper right : pseudoscalar ($\eta_b$), lower left : scalar ($\chi_{b0}$), lower right : axial-vector ($\chi_{b1}$).}
    \label{figure_finite_all}
\end{figure}

The results of the spectral functions at finite temperatures are shown in Fig. \ref{figure_finite_all}. All the channels show the same qualitative behavior. First, the peak position undergoes a shift to a lower energy with increasing temperature. Next, the peak gradually lowers, becomes broader and simultaneously shifts to slightly higher energies. At the same time, a continuum-like structure grows in the low energy region, penetrates into the peak regions and moves downward. As a further point, let us mention the crucial difference between the behavior of S-wave and P-wave channels. Quantitatively, the vector and pseudoscalar $b\bar{b}$ states, $\Upsilon$ and $\eta_b$, remain as clear peaks up to $T/T_c=2.0$ and may still survive at $T/T_c=2.5$. In the case of $\Upsilon$, one sees a bump even at $T/T_c=3.0$. On the other hand, $\chi_{b0}$ and $\chi_{b1}$ seem to disappear at $T/T_c=2.0-2.5$.

\begin{figure}[!b]
    \begin{minipage}[t]{0.5\columnwidth}
        \begin{center}
            \includegraphics[clip, width=1.0\columnwidth]{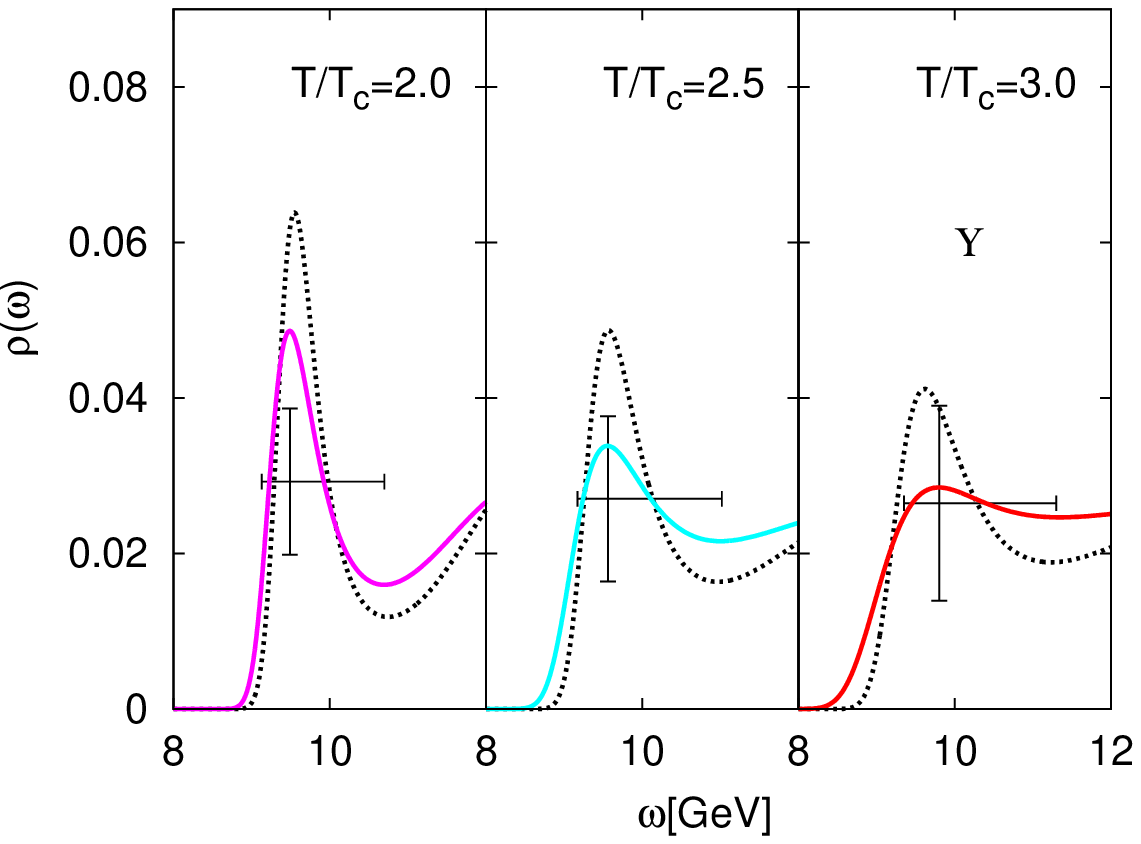}
        \end{center}
    \end{minipage}%
    \begin{minipage}[t]{0.5\columnwidth}
        \begin{center}
            \includegraphics[clip, width=1.0\columnwidth]{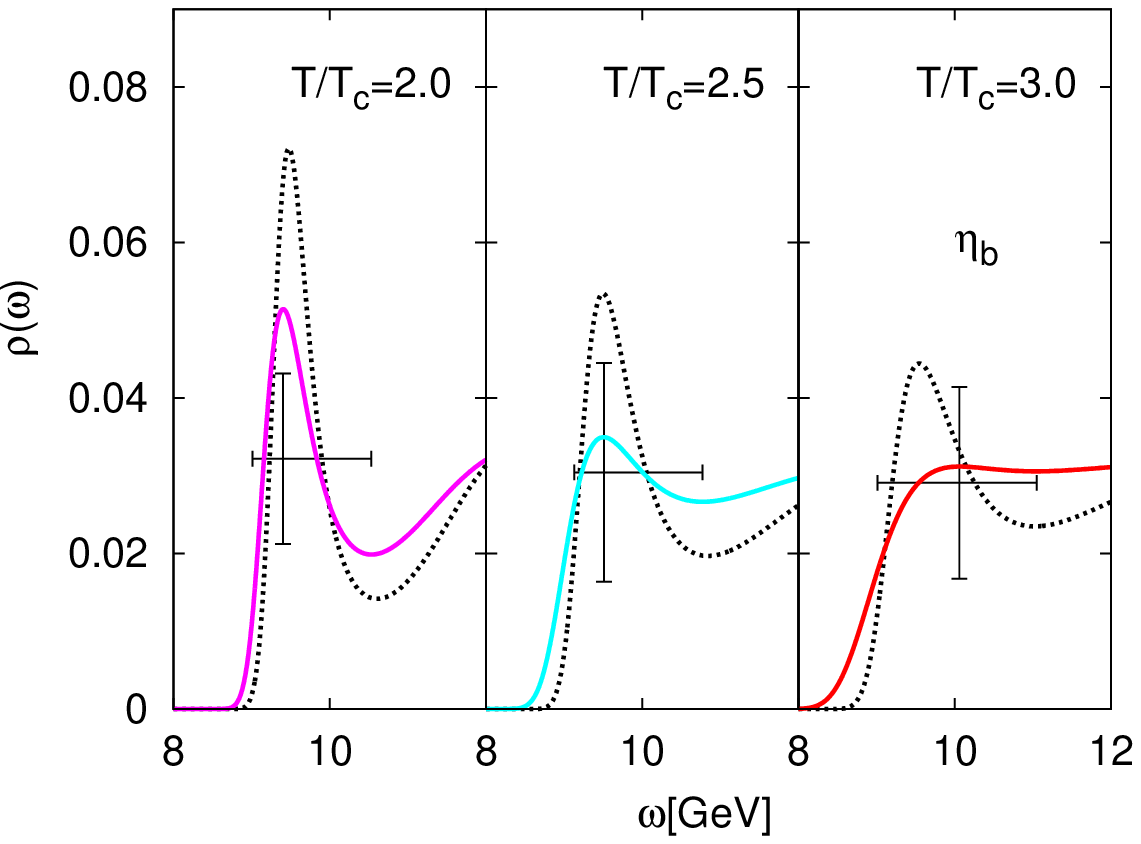}
        \end{center}
    \end{minipage}
    \begin{minipage}[t]{0.5\columnwidth}
        \begin{center}
            \includegraphics[clip, width=1.0\columnwidth]{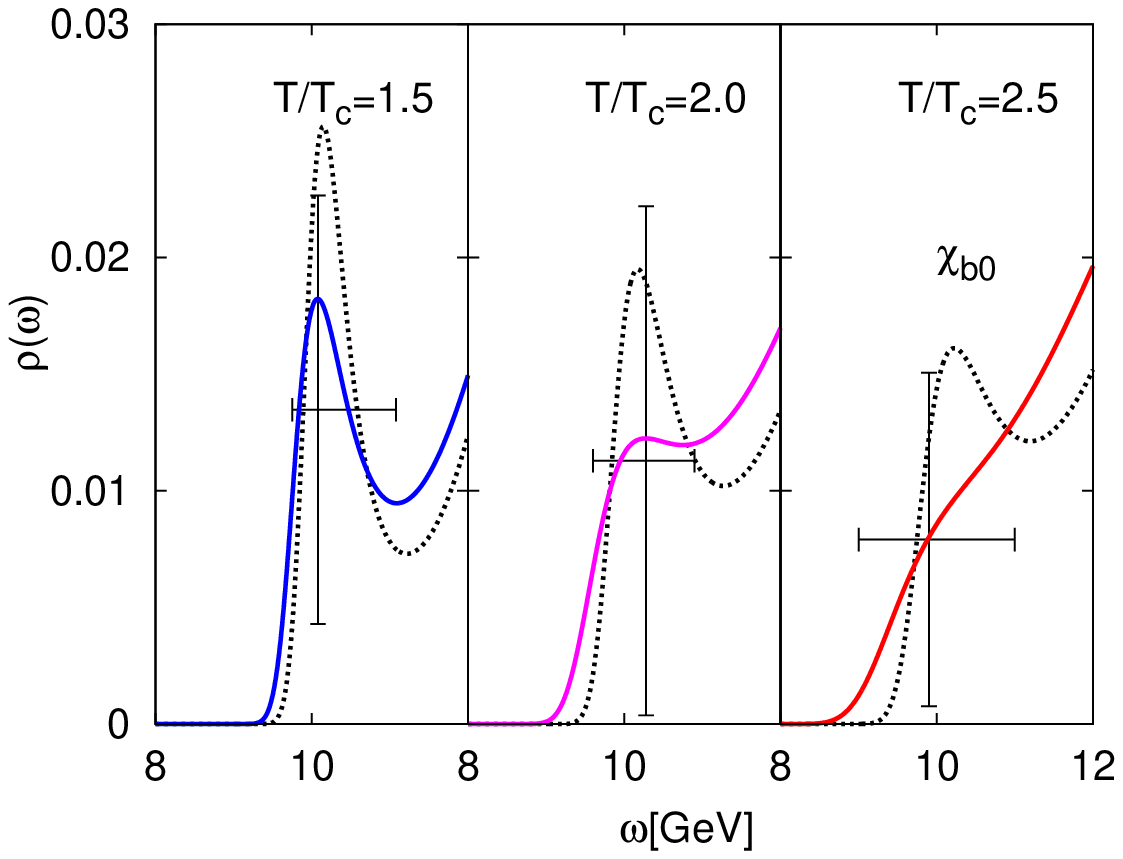}
        \end{center}
    \end{minipage}%
    \begin{minipage}[t]{0.5\columnwidth}
        \begin{center}
            \includegraphics[clip, width=1.0\columnwidth]{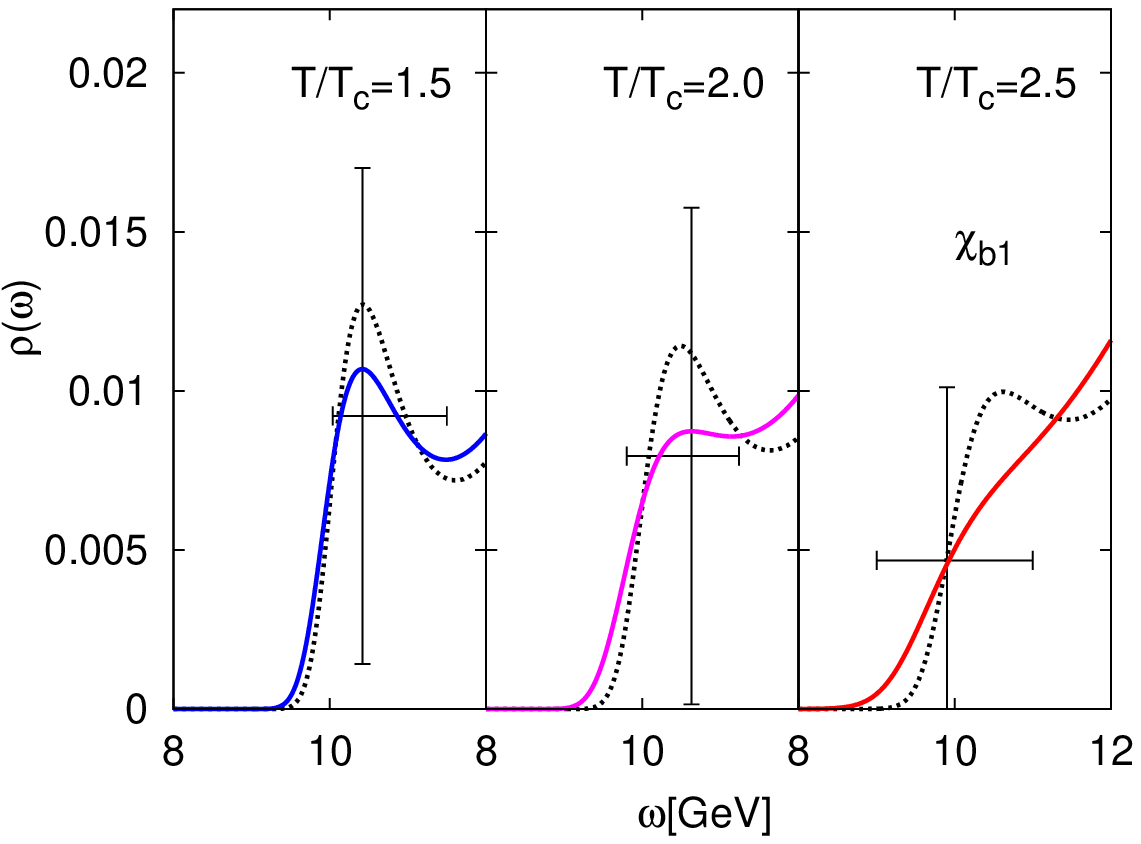}
        \end{center}
    \end{minipage}
     \caption{The solid lines and their error bars show the spectral functions of bottomonia obtained from finite temperature OPE data. The dashed lines stand for spectral functions at zero temperature using error at corresponding finite temperature. Upper left : vector ($\Upsilon$), upper right : pseudoscalar ($\eta_b$), lower left : scalar ($\chi_{b0}$), lower right : axial-vector ($\chi_{b1}$).}
     \label{figure_error}
\end{figure}

Since the uncertainties involved in the calculation are quite large, as indicated by the error bars in Fig. \ref{figure_error}, we can presently not make statements about specific numbers for the melting temperatures of the investigated states, but can only restrict the ranges of temperatures within which the peaks disappear. Concretely, we define the range of the melting temperatures as follows. The {\it upper limit} is determined as the temperature where the bump (extremum) disappears, while the {\it lower limit} taken as the temperature where the error bar exceeds the lowest-energy peak height of the spectral function, namely when the peak ceases to be statistically significant. The error bars for each temperature are shown in Fig. \ref{figure_error}. The resultant upper and lower limits of the dissociation temperatures are summarized in Table \ref{dissociation_temperature}. If we were able to calculate the OPE data with better precision, the lower (upper) limit of the temperature ranges would be increased (decreased).

\begin{table}[!t]
  \begin{center}
  \begin{tabular}{c|cccc}
\hline \hline
   Channel                                                & Vector ($\Upsilon$) & Pseudoscalar ($\eta_b$) & Scalar ($\chi_{b0}$) & Axial-vector ($\chi_{b1}$) \\
\hline
$T/T_c$                 & $> 2.3$        & $> 2.1$              & $1.3 - 2.5$ & $<2.5$                \\
\hline \hline
  \end{tabular}
  \end{center}
   \caption{Upper and lower limits of the dissociation temperature ranges for the lowest bottomonium states. The precise definition of these limits is given in the text.}
   \label{dissociation_temperature}
\end{table}

In order to confirm that the obtained results are caused by genuine physical effects, we have to check possible contributions of MEM artifacts at finite temperature. First, as the contributions of the gluon condensates increase at finite temperature, their uncertainties magnify the OPE error. Therefore, it is expected that the resolution of the MEM is reduced and the peaks of the extracted spectral functions become broader. Thus, to investigate this effect, we reanalyze the spectral functions by using the OPE at $T=0$ with error at finite temperature. The result for each channel is shown as dashed lines in Fig. \ref{figure_error}. As one can see, although the heights of the peaks are reduced partly due to the MEM artifact described above, the peaks are still present in the S-wave (P-wave) channels with the error of $T/T_c = 3.0$ ($T/T_c = 2.5)$. We also stress that the MEM artifact does not shift the peak position. From this analysis, we conclude that the disappearance of the peaks at the finite temperatures is caused by physical effects and not due to an MEM artifact.

\subsection{Excited states of bottomonia}
In Section 3.1, we showed that the spectral function extracted by the MEM contains contributions of the excited states in the lowest peak. In order to extract finite temperature effects on the excited states from the spectral functions, we analyze the residue of the lowest peak of the vector channel. However, one cannot naively integrate the spectral function in the region of the peak because the spectral function is contaminated by the continuum, which is not negligible particularly at high temperatures. 

In order to exclude the continuum contributions and to estimate the sum of the residues only of the ground and excited $b\bar{b}$ states, we fit the obtained spectral functions by using a Breit-Wigner (or Gaussian) function for the peak, and the continuum parametrized in the form of the leading order perturbative result. Specifically, we take
\begin{equation}
f(x) = \frac{|\lambda|^2}{2\pi } \frac{  \Gamma}{(x-m)^2 +\Gamma^2 /4} +\frac{1}{8\pi^2} \sqrt{1-\frac{4a^2}{x^2}} \left(2+\frac{4a^2}{x^2} \right), \label{Breit-Wigner}
\end{equation}
for fitting the MEM results. The four fitting parameters, $|\lambda|^2$, $m$, $\Gamma$, and $a$ correspond to the residue, peak position, width, and continuum threshold, respectively. Note that $a$ coincides with $m_b$ in the perturbative calculation. These parameters are fitted by the Levenberg-Marquardt method \cite{Levenberg1944,Marquardt1963}. Furthermore, in order to exclude a possible initial value dependence of the fitting procedure and to investigate the existence of local minima, we take 200 initial values generated randomly for the four fitting parameters at each temperature.

\begin{figure}[h]
    \begin{minipage}[t]{0.5\columnwidth}
        \begin{center}
            \includegraphics[clip, width=1.0\columnwidth]{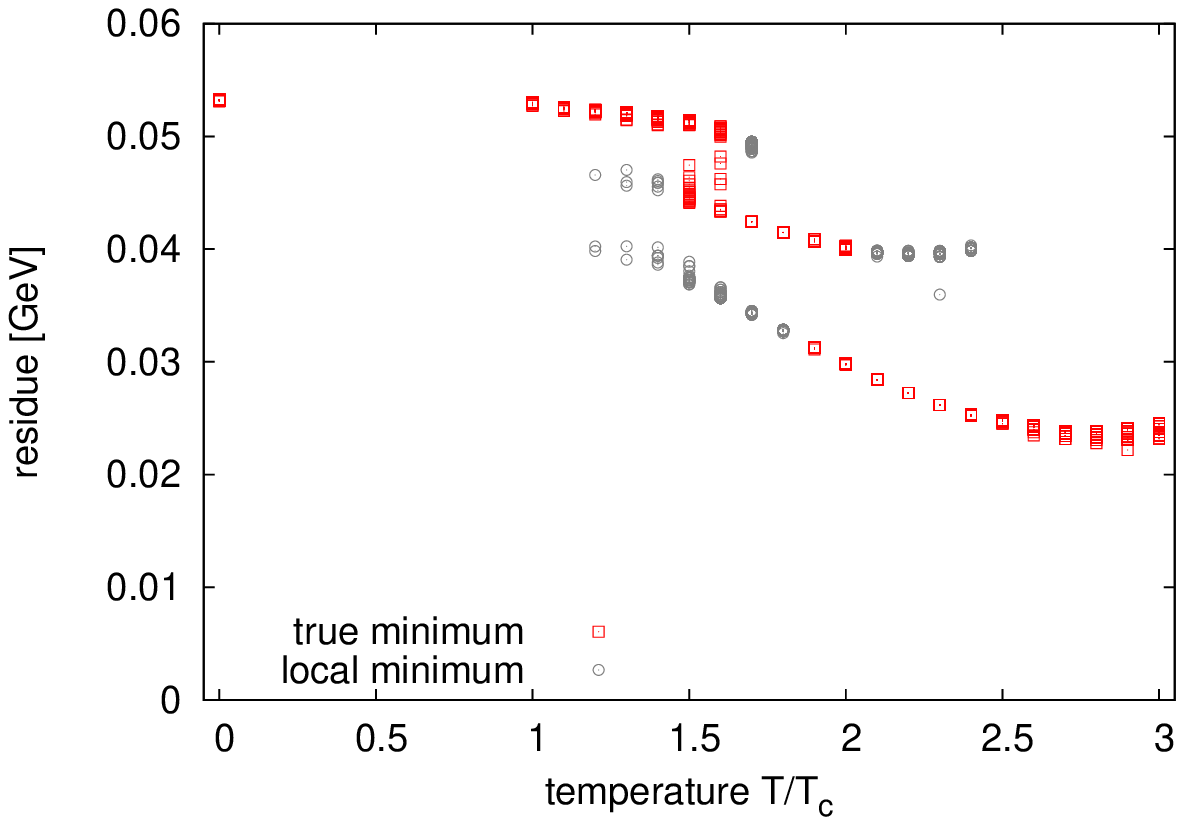}
        \end{center}
        \label{residue_BW}
    \end{minipage}%
    \begin{minipage}[t]{0.5\columnwidth}
        \begin{center}
            \includegraphics[clip, width=1.0\columnwidth]{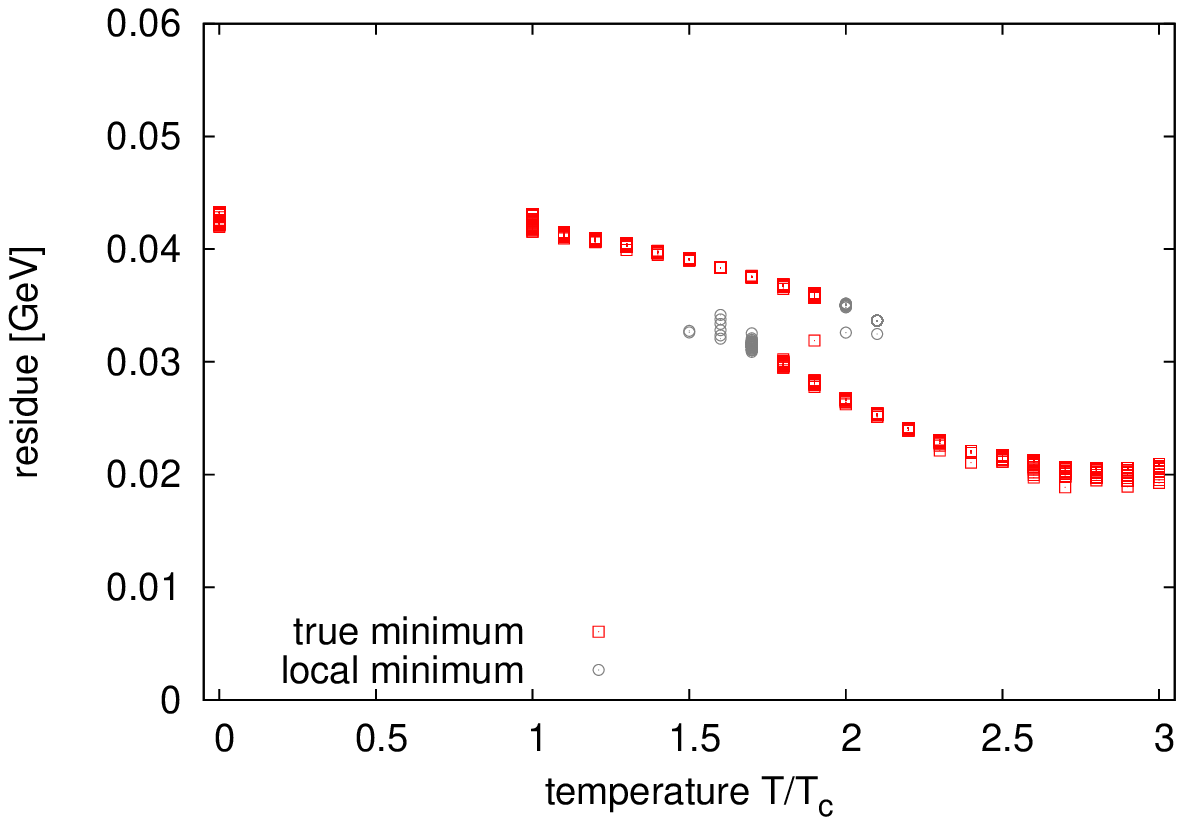}
        \end{center}
        \label{residue_GA}
    \end{minipage}
    \caption{Temperature dependence of the residue for the vector channel peak fitted with a Breit-Wigner + continuum (left) and as Gaussian + continuum (right). Red points are corresponding to true minima, and gray points stand for local minima. Fit range is fixed to 7.0-12.0 GeV.}
     \label{residue}
\end{figure}

The obtained residues of the $\Upsilon$ peak, $|\lambda|^2$, with increasing $T/T_c$ are plotted in Fig. \ref{residue}. The left panel shows the results of the fitting with the function Eq.(\ref{Breit-Wigner}). For each temperature, 200 results corresponding to initial values are plotted. For some temperature, we find multiple solutions, which are supposed to be local minima solutions of the least-square function $\chi^2$ in the L-M method. In the case of the Breit-Wigner + continuum fitting, we find that the local minimum form three clusters, top, middle, and bottom. For some $T/T_c$, one sees that the clusters are diffused and solutions are scattering to interpolate two clusters. Then, the minimum valley of $\chi^2$ seems to become flat between the two minima.

\begin{figure}[h]
    \begin{minipage}[t]{0.5\columnwidth}
        \begin{center}
            \includegraphics[clip, width=1.0\columnwidth]{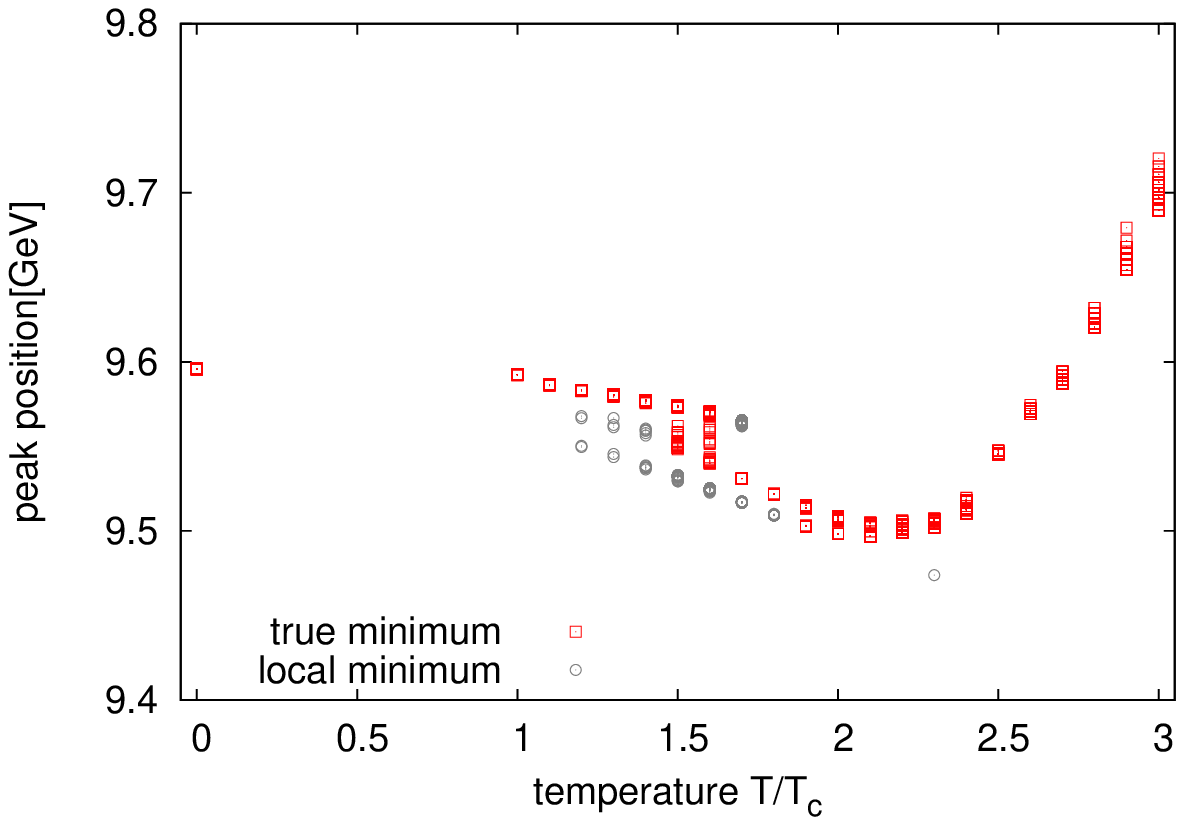}
        \end{center}
        \label{}
    \end{minipage}%
    \begin{minipage}[t]{0.5\columnwidth}
        \begin{center}
            \includegraphics[clip, width=1.0\columnwidth]{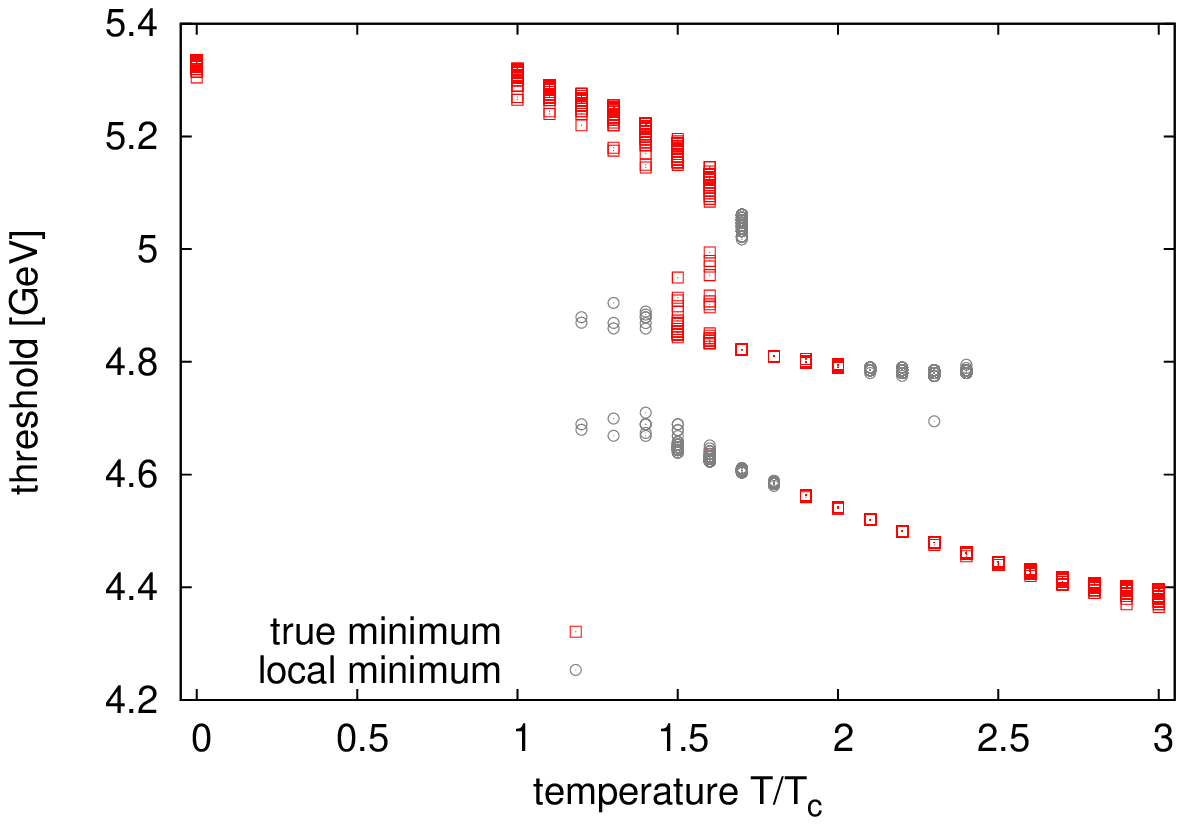}
        \end{center}
        \label{threshold_BW}
    \end{minipage}
     \caption{Temperature dependence of the fitting parameters for the vector channel peak fitted with a Breit-Wigner + continuum. Left : peak position, right : continuum threshold. Red points are corresponding to true minima, and gray points stand for local minima. Fit range is fixed to 7.0-12.0 GeV.}
     \label{mass_BW}
\end{figure}

The red point at each $T/T_c$ is the true minimum point, where $\chi^2$ hits the minimum. One sees that at low $T/T_c$, the true minimum is located in the top cluster, while it moves down to the middle cluster at around $T/T_c \sim 1.5-1.6$, and further down to the bottom cluster at around $T/T_c \sim 1.9-2.0$. At $T/T_c \geq 2.5$, we have only one stable solution. The peak positions and the continuum thresholds from the fitting with Breit-Wigner + continuum form is shown in Fig. \ref{mass_BW}. Both the peak positions and continuum threshold undergo a transition towards smaller values at $T/T_c \sim 1.5-2.0$.

To make sure that this fitting analysis is valid, we have repeated the same procedure with an alternate fitting function, namely, the Gaussian + continuum form. The results are shown in the right panel of Fig. \ref{residue}. The qualitative behaviors are the same as the previous case. It however shows only two clusters of local minima and the transition from the higher to the lower cluster at $T/T_c \sim 1.8-1.9$. Comparing these two fitting results, we conclude that the qualitative behavior of the residue is independent of fitting peak function form.

It can be concluded from the above results that the residue of $\Upsilon$ peak decreases gradually with increasing temperature and becomes a constant value at higher temperature. Especially, the rapid reduction of the residue is seen at $T/T_c = 1.5 -2.0$. It should be noted that this behavior does not directly imply that the excited states, $\Upsilon(2S)$ and $\Upsilon(3S)$, disappear at lower temperature than ground state $\Upsilon(1S)$ because one cannot eliminate other possibilities such as simultaneous reduction of the ground and excited states. Nevertheless, if we assume the disappearance of the excited states at lower temperature than the ground state, our results suggest that $\Upsilon(2S)$ and $\Upsilon(3S)$ disappear at $T/T_c = 1.5 -2.0$, while $\Upsilon(1S)$ survives up to $T/T_c = 3.0$.

\section{Conclusion} \label{Conclusion}
In summary, we have analyzed the bottomonium spectral functions at zero and finite temperature by using a newly developed analysis method of QCD sum rules. The Maximum Entropy Method (MEM) is adapted to extract the spectral function from the sum rule.

At $T = 0$, the lowest peak has been obtained for each channel corresponding to $\Upsilon$, $\eta_b$, $\chi_{b0}$, and $\chi_{b1}$. Although these mass spectra agree qualitatively with experimental values, their peak positions shift slightly to higher energies. By analyzing mock data for the vector channel and evaluating the obtained residue, we conclude that the disagreement is caused by the contribution of the excited states.

Next, we have investigated the temperature dependence of the spectral functions. Temperature dependences are taken into account in the gluon condensates, which are estimated from the quenched lattice QCD data at finite temperature. As a result, we have found that the spectral functions of bottomonia are modified much slower as functions of $T/T_c$ than those of charmonia, in which the lowest peak disappears suddenly at the vicinity of $T_c$. Using the definitions of the upper and lower limits of the melting temperature given in Section \ref{Bottomonia at finite temperature}, we find that $\Upsilon$ and $\eta_b$ survive as a peak in the spectral functions up to some temperature restricted to the regions of $T/T_c > 2.3$ and $T/T_c > 2.1$, while the dissociation temperatures of $\chi_{b0}$ and $\chi_{b1}$ are confined to $T/T_c = 1.3 - 2.5$ and $T/T_c < 2.5$, respectively.

It should be noted, however, that our definition inevitably contains some ambiguity due to the limitation of the OPE and MEM. Therefore, respective results on the melting temperature should be regarded as qualitative guides. Furthermore, both the P-wave peaks are not found to be fully significant statistically even at $T=0$, which means that we cannot make a definite conclusion about the fate of these states at finite temperature. To obtain more conclusive results on their behavior, further studies are needed once more precise information on the OPE is available. The current prediction of the melting temperatures depends on the extracted temperature dependences of the gluon condensates. For these we have used the quenched lattice QCD data for the energy density and pressure. To go beyond the quenched approximation, a more detailed analysis will be required to include full QCD information on the gluon condensates, which will be the subject of a future investigation.

Our results are qualitatively consistent with previous QCD sum rule analysis in conventional method \cite{Morita2010}. As mentioned above, however, it turns out that the lowest peaks contain the excited states as well as the ground state, so that a deformation of such a peak depends on the behavior of the excited states. Therefore, to extract more detailed information on the spectral function of the vector channel, we have investigated the temperature dependence of the residue of the lowest peak obtained from MEM for this channel. Then, we have observed that the residue decreases with increasing temperature. The results are consistent with a picture that the excited states, $\Upsilon(2S)$ and $\Upsilon(3S)$, dissociate at lower temperatures than the ground state $\Upsilon(1S)$. Our results indicate that $\Upsilon(2S)$ and $\Upsilon(3S)$ disappear in the temperature region of $T/T_c=1.5-2.0$.

\section*{Acknowledgments}
This work is partly supported by the Grant-in-Aid for Scientific Research from MEXT (No. 22105503). K.S. acknowledges the financial support from the Global Center of Excellence Program by MEXT, Japan through the ``Nanoscience and Quantum Physics'' Project of the Tokyo Institute of Technology. P.G. gratefully acknowledges the support by the Japan Society for the Promotion of Science for Young Scientists (contract No. 21.8079). K.M. is supported by Yukawa International Program for Quark-Hadron Sciences at Kyoto University (No. 24540271).





\bibliographystyle{model1a-num-names}
\bibliography{reference}







\end{document}